\documentclass[preprint,3p,times,12pt,sort&compress,a4paper]{elsarticle}

\usepackage{amssymb}
\usepackage{amsmath}
\usepackage{amsthm}

\usepackage{siunitx}
\usepackage{upgreek}
\usepackage[%
bb=px,
bbscaled=1.0,%
cal=cm,
calscaled=1.0,%
scr=boondoxo,
scrscaled=1.15,%
frak=boondox,
frakscaled=1.0,%
]{mathalpha}

\DeclareSIUnit\corehour{\text{core-hours}}
\usepackage{hyperref}
\usepackage[nameinlink,capitalize]{cleveref}
\usepackage{enumitem}

\usepackage{lineno}

\journal{Nucl. Instrum. Methods Phys. Res.}

\begin{document}

\begin{frontmatter}

\title{Development and validation of a high-fidelity full-spectrum Monte Carlo model for the Swiss airborne gamma-ray spectrometry system}

\author[1,2]{David Breitenmoser\corref{cor1}}
\ead{david.breitenmoser@psi.ch}
\author[1]{Alberto Stabilini}
\author[1]{Malgorzata Magdalena Kasprzak}
\author[1]{Sabine Mayer}

\affiliation[1]{organization={Department of Radiation Safety and Security, Paul~Scherrer~Institute (PSI)},
            addressline={Forschungsstrasse~111}, 
            city={\\Villigen~PSI},
            postcode={5232}, 
            country={Switzerland}}
\affiliation[2]{organization={Department of Physics, Swiss Federal Institute of Technology (ETH)},
            addressline={Otto-Stern-Weg~5}, 
            city={\\Zurich},
            postcode={8093}, 
            country={Switzerland}}
\cortext[cor1]{Corresponding author.}

\begin{abstract}
Airborne Gamma-Ray Spectrometry (AGRS) is a critical tool for radiological emergency response, enabling the rapid identification and quantification of hazardous terrestrial radionuclides over large areas. However, existing calibration methods are limited to only a few gamma-ray sources, excluding most radionuclides released in severe nuclear accidents and nuclear weapon detonations, which compromises effective response and accurate risk assessments in such incidents. Here, we present a high-fidelity Monte Carlo model that overcomes these limitations, offering full-spectrum calibration for any gamma-ray source. Unlike previous approaches, our model integrates a detailed mass model of the aircraft and a calibrated non-proportional scintillation model, enabling accurate event-by-event predictions of the spectrometer's response to arbitrarily complex gamma-ray fields. Validation measurements in near-, mid-, and far-field scenarios demonstrate that the developed model not only effectively addresses the large deficiencies of previous approaches, but also achieves the accuracy and precision required to supersede traditional empirical calibration methods. These results mark a major advancement in AGRS. The developed calibration methodology not only allows for the generation of high-fidelity spectral signatures for any gamma-ray source, but also reduces calibration time and costs, minimizes reliance on high-intensity calibration sources, and eliminates the generation of related radioactive waste. The approach presented here serves as a critical step toward integrating advanced full-spectrum data reduction methods for AGRS, which could unlock promising new capabilities for these systems beyond emergency response, including the quantification of the cosmic-ray flux in the atmosphere for geophysical research and the identification of trace-level airborne radionuclides in nuclear security applications.

\end{abstract}


\end{frontmatter}

\section{Introduction}
\label{sec:Introduction}


\noindent By mounting high-volume gamma-ray spectrometers in an aircraft, Airborne Gamma-Ray Spectrometry (AGRS) systems enable the localization, identification, and quantification of terrestrial gamma-ray sources over extensive areas \mbox{\ensuremath{\mathcal{O}(\num{1d8})\,\unit{\square\m}}} within \mbox{\ensuremath{\mathcal{O}(\num{1d2})\,\unit{\minute}}} \citep{Breitenmoser2024a}. This unique capability makes AGRS essential for a wide range of applications, including radiological emergency response, geological mapping, mineral exploration, and environmental monitoring \citep{Connor2016a,Li2020,PradeepKumar2020}.

The accuracy and sensitivity of AGRS systems rely heavily on the adopted calibration protocols. The International Atomic Energy Agency (IAEA) has standardized these methods in two technical reports \citep{IAEA2003,IAEA1991}, which utilize simplified physics models combined with empirical calibration. These methods, while useful, have critical limitations.

Current empirical calibration methods are restricted to a limited range of radionuclide sources and source-detector configurations. Comprehensive calibration can require up to \qty{5}{\day} to cover all relevant source-detector configurations and often involves high costs for source preparation \mbox{and/or} decommissioning \citep{Dickson1981,Grasty1991,Minty1990}. Additionally, these methods are subject to significant systematic uncertainties, primarily originating from varying radiation backgrounds and required analytical corrections.

The physics models employed, on the other hand, are only reliable for photon energies \mbox{\ensuremath{\gtrsim\!\qty{400}{\keV}}} \citep{IAEA2003,IAEA1991}. Moreover, these models are limited to narrow spectral bands, which greatly reduces the AGRS sensitivity by excluding much of the available information encoded in the pulse-height spectra \citep{Hendriks2001,Grasty1985}.

These limitations in current AGRS calibration protocols significantly compromise the quality and reliability of the derived data for emergency response applications. This became particularly evident in the severe nuclear accident at the Fukushima Daiichi nuclear power plant in 2011, where the current calibration limitations in AGRS severely hindered effective response and accurate risk assessment \citep{Lyons2012}. Improving the current calibration and data evaluation protocols in AGRS to address the discussed limitations in the capabilities of AGRS systems is therefore of immediate relevance for public safety.

To overcome the discussed limitations in AGRS calibration and data evaluation, prior studies have proposed following a numerical instead of an empirical approach, specifically high-fidelity Monte Carlo simulations \citep{Allyson1998,Billings1999}. Monte Carlo models offer the potential to predict the full-spectrum gamma-ray spectrometer response without restrictions in the source type or source-detector geometry \citep{Ahdida2022,Allison2016,Goorley2016,Sato2024}. However, implementing Monte Carlo based calibration and data evaluation for AGRS has proven challenging due to the model's complexity and high computational cost \citep{Allyson1998,Billings1999,Sinclair2011,Sinclair2016,Zhang2018,Kulisek2018,Torii2013}. These difficulties arise mainly from the large simulation domain \mbox{\ensuremath{\mathcal{O}(\num{1d10})\,\unit{\cubic\m}}} and the modeling effort required to develop a sufficiently accurate mass model of the aircraft platform with characteristic lengths of \mbox{\ensuremath{\mathcal{O}(\num{1d1})\,\unit{\m}}}. 

To reduce model complexity and computational costs, earlier studies either omitted the aircraft entirely or represented it as primitive solids in their Monte Carlo models \citep{Allyson1998,Billings1999,Sinclair2011,Torii2013,Sinclair2016,Zhang2018,Kulisek2018}. These simplifications have proven to introduce significant systematic errors in the simulations, in particular at photon energies \mbox{$\lesssim\!\qty{1d2}{\keV}$} with relative errors exceeding \mbox{\ensuremath{\qty{200}{\percent}}} \citep{Allyson1998,Kulisek2018}. Moreover, most studies followed traditional IAEA methods, focusing on narrow spectral bands through a hybrid approach that integrates monoenergetic flux models with Monte Carlo simulations, rather than predicting the full-spectrum response of the gamma-ray spectrometer \citep{Billings1999,Sinclair2011,Torii2013,Sinclair2016,Zhang2018}. As a result, Monte Carlo based methods have not yet reached the accuracy or sensitivity required to supersede traditional AGRS methodologies, leaving the fundamental limitations in AGRS discussed above unresolved.

The main scope of this work is the development and validation of an advanced full-spectrum Monte Carlo model for the Swiss AGRS system, aiming to overcome the fundamental limitations of traditional AGRS calibration methods, particularly their restriction to a limited range of radionuclide sources, source-detector configurations, and gamma-ray energies. The developed Monte Carlo model features two key innovations: (1)~a high-fidelity mass model of the entire aircraft to reduce systematic errors noted in previous studies, and (2)~the integration of an advanced non-proportional scintillation model (NPSM) that enables accurate event-by-event prediction of the full-spectrum response of inorganic scintillators. Together, these advancements allow us to perform full-spectrum calibration of the Swiss AGRS system for both point and extended gamma-ray sources with an unprecedented level of accuracy. We demonstrate this through a series of validation measurements conducted under various field conditions with both natural and man-made radionuclides. The validation of the gamma-ray spectrometer mass model under laboratory conditions has been presented in earlier studies \citep{Breitenmoser2023c,Breitenmoser2022}.

\section{Methods}
\label{sec:Methods}

\subsection{Swiss AGRS system}
\label{sub:SwissAGRS}

\noindent Here, we review the technical specifications of Switzerland’s currently operational AGRS system, Radiometrie Land-Luft (RLL), hereafter referred to as the Swiss AGRS system. A comprehensive discussion of its development, operational capabilities, and integration within Swiss civil and military organizations is provided by \citet{Breitenmoser2024a}.

\begin{figure}[t]
\centering
\includegraphics[width=1\textwidth]{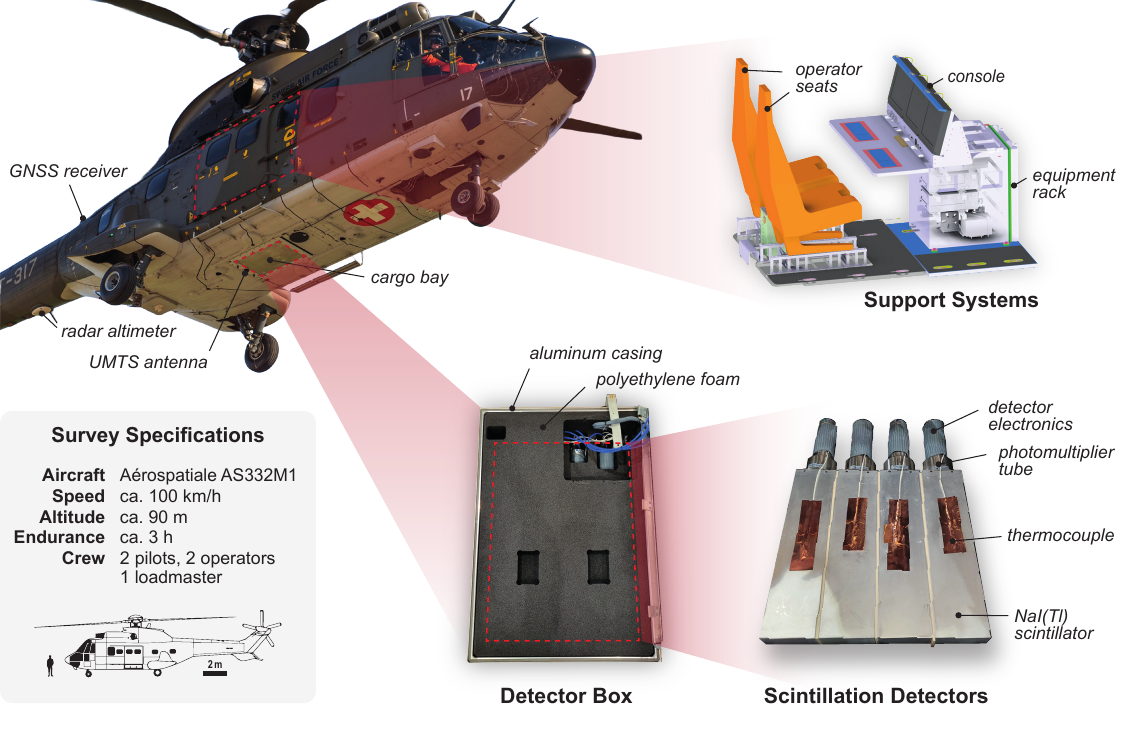}
\caption[Overview of the Swiss AGRS system]{Overview of the Swiss AGRS system. This illustration highlights a close-up view of the spectrometer location within the fuselage of the AS332M1 helicopter, the detector box, the scintillation detectors as well as the support systems. In addition, selected secondary helicopter sensors accessible to the AGRS system, i.e.~global navigation satellite system (GNSS) receiver and radar altimeter, as well as the universal mobile telecommunications system (UMTS) antenna are marked. The line drawing of the AS332M1 helicopter was adapted from Jetijones, \href{https://creativecommons.org/licenses/by/3.0}{\texttt{CC~BY~3.0}}, via Wikimedia Commons.}
\label{fig:SystemOverview}
\end{figure}

As illustrated in \cref{fig:SystemOverview}, the Swiss AGRS system comprises four identical $\qty{10.2}{\cm}\times\qty{10.2}{\cm}\times\qty{40.6}{\cm}$ prismatic NaI(Tl) scintillation crystals (Saint-Gobain 4*4H16/3.5-X) protected by individual aluminum casings. Each crystal is coupled to a separate photomultiplier tube, Hamamatsu R10755, with associated detector electronic components (preamplifier and multichannel analyzer) allowing individual read-out of each scintillation crystal. The four scintillation spectrometer assemblies are embedded in a thermal-insulating and vibration-damping polyethylene foam, further protected by a rugged aluminum box with outer dimensions of $\qty{90}{\cm}\times\qty{64}{\cm}\times\qty{35}{\cm}$. The total mass of the detector box is \qty{\sim90}{\kg}. 

The spectrometer adopts a bin-mode data acquisition scheme with a sampling time of \qty{1}{\s} for each of the four scintillation crystals. These four acquisition channels will be referred to as detector channels \texttt{\#DET1} through \texttt{\#DET4} in this study. In addition to these four single channels, there is an additional fifth channel, \texttt{\#DET5}, which is simply the sum of the channels \texttt{\#DET1} through \texttt{\#DET4}. All five detector channels comprise \num{1024} pulse-height channels, with the gain and lower-level discriminator (LLD) configured to encompass a spectral range between \qty{\sim30}{\keV} and \qty{\sim3.072}{\MeV}. Furthermore, the spectrometer features automatic gain stabilization, spectrum linearization with offset correction as well as automatic live time recording, significantly simplifying the data postprocessing as discussed by \citet{Breitenmoser2022}. In addition to the primary gamma-ray spectrometer, the system is also equipped with a Geiger-M\"{u}ller tube, Centronic ZP1202, for high dose-rate measurements up to \qty{40}{\milli\sievert\per\hour}. 

Before deployment, the detector box is mounted in the cargo bay of an A\'{e}rospatiale AS332M1 Super Puma helicopter, with its bottom aligned to the helicopter's underside for optimal sensitivity to terrestrial gamma-ray sources, as illustrated in \cref{fig:SystemOverview}. A rugged computer functions as the system's data server, while two additional rugged and redundant client computers serve as interfaces for online evaluation, data mapping and communication. These computers are housed within an equipment rack situated inside the crew cabin. The total mass of the system components within the crew cabin is \qty{\sim290}{\kg}. Using an internal ARINC~429 avionics data bus, the data server computer accesses secondary aircraft sensor data, including air pressure and air temperature sensors, global navigation satellite system receiver, and radar altimeter. The system also includes a universal mobile telecommunications system (UMTS) modem and related antenna for real-time data transmission to the ground station. A schematic overview of the Swiss AGRS's system architecture is provided in Supplementary Fig.~S4. 

During surveys, two crew members, referred to herein as operators, control the system with their associated client computers, displays, keyboards, and trackballs, as highlighted in \cref{fig:SystemOverview}. Data acquisition during the survey flights is performed using a proprietary software suite (\texttt{SpirIDENT}) developed and maintained by Mirion Technologies. A detailed description of the code is provided by \citet{Butterweck2018b}.

\subsection{Monte Carlo model}
\label{sub:MCmodel}

\noindent We adopted the \texttt{FLUKA} code, Version \texttt{4-2.2}, maintained by the FLUKA.CERN Collaboration \citep{Bohlen2014,Battistoni2015,Ahdida2022} together with the graphical interface \texttt{FLAIR} \citep{Vlachoudis2009}, Version \texttt{3.2-4.5}, to perform the Monte Carlo simulations presented in this study. All Monte Carlo simulations were conducted on a local computer cluster, featuring \num{520} cores at a nominal clock speed of \qty{2.6}{\giga\Hz}.

Following a bottom-up modeling approach, we integrated the already validated non-proportional scintillation Monte Carlo (NPSMC) model of the spectrometer presented in previous studies \citep{Breitenmoser2023c,Breitenmoser2022}. Consequently, the derivation of the Monte Carlo model for the Swiss AGRS system primarily involved extending the mass model of the spectrometer to include the aircraft, while all physics models and postprocessing routines remained unchanged. In this section, we limit the discussion therefore to the extension of the mass model. Comprehensive information on the physics models, scoring, and postprocessing routines is available in \citep{Breitenmoser2023c,Breitenmoser2022}.

To incorporate the aircraft system components into the already validated spectrometer mass model from \citet{Breitenmoser2023c}, we used \texttt{FLAIR}'s and \texttt{FLUKA}'s core combinatorial solid geometry (csg) builder \citep{Vlachoudis2009,Ahdida2022}. Geometrical and material properties of the various aircraft components were obtained from a combination of sources, including expert interviews, technical schematics, and computer-aided design (CAD) files. Due to their significant size and mass, the support systems installed in the crew cabin during survey flights were also included in the mass model to ensure accurate representation of the entire AGRS system.

\begin{figure}[t!]
\centering
\includegraphics[scale=0.99]{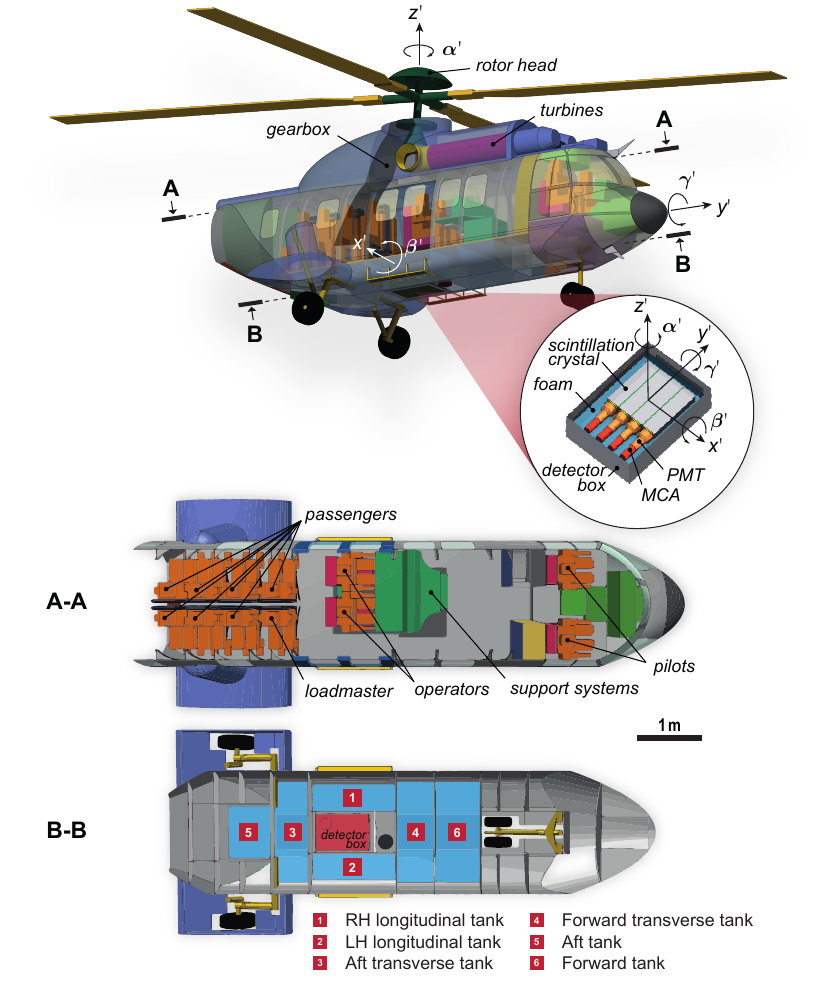}
\caption[Monte Carlo mass model]{Monte Carlo mass model. Here, we illustrate the derived Monte Carlo mass model of the Swiss AGRS system including the AS332M1 aircraft and the spectrometer system. For reference, we also indicate the adopted principal axes $x'$-$y'$-$z'$ and corresponding Tait-Bryan angles, i.e.~yaw (\ensuremath{\alpha'}), pitch (\ensuremath{\beta'}), and roll (\ensuremath{\gamma'}) quantifying the orientation of the mass model with respect to the world frame. Note that for visualization purposes, the rear of the cabin was shortened. A-A Cross-section view of the aircraft cockpit and the cabin including the support systems and the crew (max. passenger capacity). B-B Cross-section view of the fuselage highlighting the position of the six fuel tanks relative to the detector box (landing gear retracted). All mass model figures displayed here were created using the graphical interface FLAIR \citep{Vlachoudis2009}. For better visibility and interpretability, semi-transparent false colors were applied.}
\label{fig:ModelOverview}
\end{figure}

Given the complexity of the individual subsystems, the mass model was built in a modular fashion, with the individual components being incorporated as separate homogeneous objects. For these simplifications, care was taken to preserve the overall geometry, the opacity as well as the mass density. Complex geometries of some aircraft components like the fuselage were modeled by fitting quadrics to the corresponding aircraft surfaces. The resulting mass model of the AGRS detector system, featuring over \num{1d3} bodies and \num{1d2} different materials, is illustrated in \cref{fig:ModelOverview}. 

A key consideration in the derivation of AGRS mass models is the dynamic nature of the aircraft system. Previous studies discussed in \cref{sec:Introduction} either completely neglected the aircraft system or represented it as a static object using primitive geometric shapes. However, aircraft systems employed in AGRS are inherently dynamic, with many components changing position, composition, and orientation during or between survey flights. Some of this dynamic behavior is expected to have a significant impact on the measured pulse-height spectra. Therefore, in contrast to previous studies, we explicitly accounted for the following dynamic changes in the aircraft system:

\begin{enumerate}[label=\textbf{\arabic*}.]
    \item \textbf{Fuel} The AS332M1 helicopter possesses six fuel tanks with a total capacity of \qty{\sim2}{\cubic\m} holding \qty{\sim1.6d3}{\kg} of jet fuel A-1. As illustrated in \cref{fig:ModelOverview}, these tanks are arranged around the spectrometer within the helicopter's fuselage. To maintain the aircraft's center of gravity, the six fuel tanks are emptied in a predetermined sequence during the course of a survey flight (cf. Supplementary Fig.~S5). To account for these changes, we modeled the fuel level in each tank separately as a dynamic parameter following the aircraft's fuel depletion during the course of a survey flight.

    \item \textbf{Crew} The core personnel for the Swiss AGRS system consists of two pilots, operating the aircraft in the cockpit in the front, as well as a loadmaster and two operators situated in the cabin. In addition to this core crew, a variable number of up to seven passengers can be seated in the rear of the aircraft. An illustration of the cockpit and cabin layout is provided in \cref{fig:ModelOverview}. To account for changes in the number of passengers across several flights, the presence or absence of the individual passengers is treated as a dynamic parameter within the derived AGRS mass model. The crew members were modeled using mass, elemental composition, and anthropometric dimensional data for a reference human male provided by the International Commission on Radiological Protection (ICRP) \citep{Snyder1975b} and the National Aeronautics and Space Administration (NASA) \citep{Whitmore2012a}.

    \item \textbf{Orientation} In order to account for the orientation of the aircraft system with respect to the world frame, we utilized rotation transformations cards (\texttt{ROT-DEFIni}) implemented in \texttt{FLUKA}. For this purpose, three intrinsic rotations are applied sequentially about the principal axes of the aircraft in the order $z'$-$x'$-$y'$, with the corresponding Tait-Bryan angles being denoted by \ensuremath{\alpha'} (yaw), \ensuremath{\beta'} (pitch), and \ensuremath{\gamma'} (roll), as indicated in \cref{fig:ModelOverview}. It is important to note that there is no universally accepted convention for orienting the principal axes in Tait-Bryan rotations. In this work, the aircraft's principal axes are defined as follows: $z'$ points upward parallel to the fuselage station, $y'$ aligns with the longitudinal axis pointing forward and $x'$ aligns with the transverse axis pointing to starboard. The origin of the aircraft reference system is defined as the center of gravity of the four scintillation crystals.
\end{enumerate}

\subsection{Model validation}
\label{sub:ModelValidation}

\noindent To validate the Monte Carlo model described above, we performed extensive radiation measurements with the fully integrated Swiss AGRS system under near-, mid-, and far-field conditions. Postprocessing of the measured pulse-height spectra was performed using the \texttt{RLLSpec} pipeline described in Supplementary Method~S1.1, along with a thorough assessment of statistical and systematic uncertainties detailed in Supplementary Method~S1.4.1. The spectral energy, the spectral resolution as well as the LLD were calibrated using the \texttt{RLLCal} pipeline detailed in Supplementary Method~S1.2. The calibration of the non-proportional scintillation model via Compton edge probing was previously completed and is described in detail by \citet{Breitenmoser2023c}. Given the extensive set of measurements and simulations performed, this study focuses on the acquisition channel \texttt{\#DET5}, the primary channel used in current postprocessing pipelines for evaluating Swiss AGRS survey flights \citep{Butterweck2018b}. The following subsections detail the experimental setups and measurements performed under each field condition, as well as the corresponding Monte Carlo simulations.

\subsubsection{Near-field measurements}
\label{subsub:NearFieldMethod}

\begin{figure}[tb]
\centering
\includegraphics[scale=1]{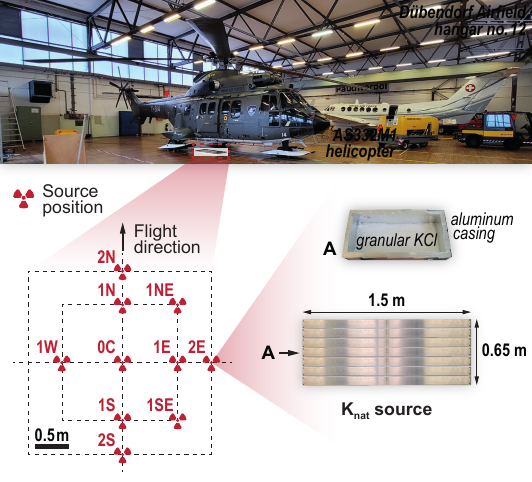}
\caption[Experimental setup of the near-field radiation measurements]{Experimental setup of the near-field radiation measurements. The main elements of the setup are the fully integrated Swiss AGRS system located in Hangar No.~12 of the Dübendorf Airfield (\qty{47.405}{\degree}N, \qty{8.643}{\degree}E) and the custom-made $\text{K}_{\text{nat}}$ radionuclide sources consisting of granular KCl sealed into aluminum casings. These sources were placed on ten different predefined positions on the hangar floor covering an area of \qtyproduct{1.5 x 0.65}{\m}. Note that for visualization purposes, the top cover of the aluminum casing in the zoomed-in view of the $\text{K}_{\text{nat}}$ radionuclide source is removed.}
\label{fig:NearFieldSetup}
\end{figure}

\noindent The aim of the near-field radiation measurements was to validate the spectral and angular accuracy of the Monte Carlo model discussed in the previous section. These measurements focused on extended radionuclide sources positioned beneath the AS332M1 aircraft, where the source-detector distance was small compared to the aircraft's characteristic dimensions (\qty{\sim10}{\m}). The radiation measurements were carried out at the Dübendorf Airfield (ICAO airport code: LSMD), specifically in Hangar No.~12 of the Swiss Air Force (\ang{47.405}N, \ang{8.643}E), over a four-day period between December 14 and 17, 2021. As illustrated in \cref{fig:NearFieldSetup}, the measurements were performed with the fully integrated Swiss AGRS system parked inside Hangar No.~12.

Custom-built $\text{K}_{\text{nat}}$ radionuclide sources, consisting of granular KCl sealed in aluminum casings, were used for the radiation measurements (see \cref{fig:NearFieldSetup}). In total, seven sources have been employed for each measurement with a total activity $\mathcal{A} = \qty{456.6(2.6)}{\kilo\becquerel}$. Details on the KCl mass and corresponding activities are provided in Supplementary Table~S1. A technical drawing of the sources is available in Supplementary Fig.~S6. If not otherwise noted, we report uncertainties herein as 1~standard deviation values in least significant figure notation. Rounding of uncertainty values is performed according to the rounding rules established by the Particle Data Group \citep{ParticleDataGroup2022}.

In total, ten radiation measurements were performed. For each measurement, the seven $\text{K}_{\text{nat}}$ radionuclide sources were arranged on predefined positions on the hangar floor, covering an area of $\qty{1.5}{\m}\times\qty{0.65}{\m}$, as indicated in \cref{fig:NearFieldSetup}. Source positions spanned polar angles $\theta'$ from $\qty{110}{\degree}$ (2E, 2N, 2S) to $\qty{180}{\degree}$ (0C), relative to the principal axis $z'$ (see \cref{fig:ModelOverview}). To arrange the sources accurately, distance as well as alignment laser systems were used. Between the gross radiation measurements, background measurements were performed regularly for background correction and gain stability checks. Measurement live times for both gross and background measurements are provided in Supplementary Table~S2.

\begin{figure}[tb]
\centering
\includegraphics[scale=1]{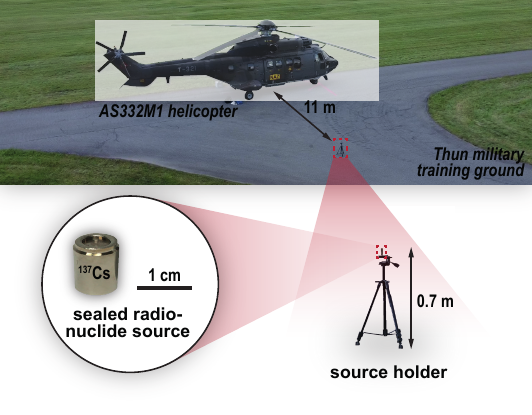}
\caption{Experimental setup of the mid-field radiation measurements. The main elements of the setup are the fully integrated Swiss AGRS system parked on the ground at the Thun military training ground (\qty{46.753}{\degree}N, \qty{7.596}{\degree}E) and a sealed \mbox{$^{137}\text{Cs}$} radionuclide point source ($\mathcal{A} = \qty{7.7(4)d8}{\becquerel}$). This point source was inserted in a custom source holder made from polylactide polymer (PLA), mounted on a tripod and positioned along the principal axis $x'$ on the starboard side of the AS332M1 helicopter in a distance of \qty{11.0(5)}{\m} with respect to the center of gravity of the scintillation crystals. Note that the adopted zoom-in figure of the radionuclide source is only representative of its size and geometry and does not depict the actual source deployed during the measurements.}
\label{fig:MidFieldSetup}
\end{figure}

\subsubsection{Mid-field measurements}
\label{subsub:MidFieldMethod}

\noindent The mid-field radiation measurements aimed to validate the Monte Carlo model under conditions where the aircraft’s fuselage significantly attenuated the primary gamma-ray field and with the source-detector distance being comparable to the aircraft's characteristic dimensions (\qty{\sim10}{\m}). The measurements were conducted at the Thun military training ground (\qty{46.753}{\degree}N, \qty{7.596}{\degree}E) on June 16, 2022, during the civil part of the ARM22 exercise (ARM22c) \citep{Butterweck2023a}. For these measurements, the AS332M1 helicopter was stationed on the ground at the training site, as shown in \cref{fig:MidFieldSetup}. After the helicopter was parked, we placed a sealed \mbox{$^{137}\text{Cs}$} point source ($\mathcal{A} = \qty{7.7(4)d8}{\becquerel}$) at a distance of \qty{11}{\m} with respect to the center of gravity of the scintillation crystals along the principal axis $x'$ on the starboard side of the AS332M1 helicopter. The source was mounted on a tripod with a custom polylactide polymer (PLA) source holder fabricated using additive manufacturing. To place the sources accurately, we used distance as well as alignment laser systems. The source’s placement relative to the fuel tanks ensured that the line of sight to the detector was obstructed by both the helicopter fuselage and fuel tank \#2 (cf. \cref{fig:ModelOverview}). This configuration allowed us to validate both the Monte Carlo mass model representation of the fuselage and the dynamic fuel level model detailed in \cref{sub:MCmodel}.

To assess the effect of the aircraft fuel on the detector response in the given source-detector configuration, measurements were repeated three times with three different fuel volume fractions ($\varrho_{\scriptscriptstyle\mathrm{JF}}$): \qty{9.5}{\percent}, \qty{43.7}{\percent}, and \qty{92.2}{\percent}. For that purpose, the helicopter was gradually refueled in the field between the individual measurements using a tanker truck. To correct for background radiation, additional background measurements were conducted for each gross measurement using the same source-detector configuration but with the source removed. A list of all performed measurements with associated gross and background measurement live times is provided in Supplementary Table~S3.

\subsubsection{Far-field measurements}
\label{subsub:FarFieldMethod}

\begin{figure}[tb]
\centering
\includegraphics[scale=1]{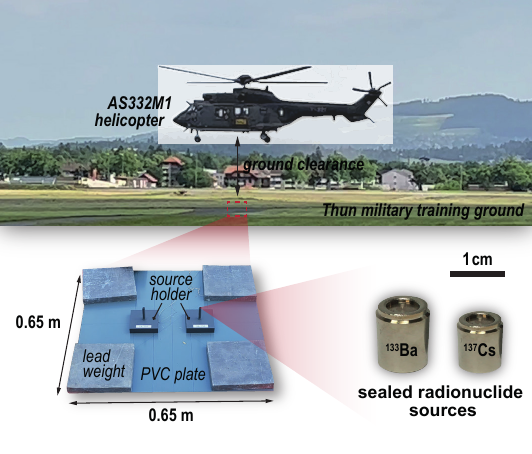}
\caption{Experimental setup of the far-field radiation measurements. The main elements of the setup are the fully integrated Swiss AGRS system operated in hover flight mode at a predefined ground clearance above a sealed radionuclide point source positioned on the ground in a custom source holder. Two different sources have been deployed separately: a \mbox{$^{137}\text{Cs}$} source ($\mathcal{A} = \qty{9.0(5)d9}{\becquerel}$) and a \mbox{$^{133}\text{Ba}$} source ($\mathcal{A} = \qty{4.7(2)d8}{\becquerel}$). The source holder consisted of a polyvinyl chloride (PVC) base plate with two custom source holders made from polylactide polymer (PLA) and constructed using additive manufacturing techniques. The base plate was secured to the ground using four lead bricks, each with a mass of \qty{11.3}{\kg}. Note that the adopted zoom-in figures of the radionuclide sources are only representative of their size and geometry and do not depict the actual sources deployed during the measurements.}
\label{fig:FarFieldSetup}
\end{figure}

\noindent The far-field radiation measurements were designed to validate the Monte Carlo model under realistic source-detector conditions representative of AGRS surveys, where the source-detector distance is significantly larger than the characteristic dimensions of the aircraft (\qty{\sim10}{\m}). At these distances, environmental factors, particularly atmospheric effects, have an increasingly pronounced influence on the detector response. Consequently, these measurements provided a basis for validating not only the mass model of the aircraft and detector system but also the environmental model.

The far-field measurements, like the mid-field ones, were conducted at the Thun military training ground (\qty{46.753}{\degree}N, \qty{7.596}{\degree}E) on June 16, 2022, during the civil part of the ARM22 exercise (ARM22c) \citep{Butterweck2023a}. In these radiation measurements, the Swiss AGRS system was operated in a hover flight mode with the AS332M1 helicopter positioned at a predefined ground clearance above a sealed radionuclide point source placed on the ground in a custom source holder aligned with the principal axis $z'$, as indicated in \cref{fig:FarFieldSetup}. The hover flight was performed at three different constant ground clearance levels: \qty{\sim30}{\m}, \qty{\sim60}{\m}, and \qty{\sim90}{\m} with a measurement live time of \qty{\sim5}{\minute} for each measurement. As illustrated in \cref{fig:FarFieldSetup}, the source holder consisted of a polyvinyl chloride (PVC) base plate with two custom source holders made from PLA and constructed using additive manufacturing techniques. The base plate was secured to the ground using four lead bricks, each with a mass of \qty{11.3}{\kg}.

Two sealed radionuclide sources, a \mbox{$^{137}\text{Cs}$} source ($\mathcal{A} = \qty{9.0(5)d9}{\becquerel}$) and a \mbox{$^{133}\text{Ba}$} source ($\mathcal{A} = \qty{4.7(2)d8}{\becquerel}$), were deployed separately for each hover flight by placing them in the corresponding source holder. To correct for the background radiation, additional background measurements were performed for each gross measurement using the same source-detector configuration but with the sources removed.

Ground clearances were determined using the radar altimeter of the AS332M1 helicopter accessed by the ARINC~429 avionics data bus. In addition, orientation data of the helicopter was obtained by the flight data recorder. A list of all performed measurements with associated gross and background measurement live times, mean ground clearances, and orientation angles of the helicopter is provided in Supplementary Table~S4.

\subsubsection{Monte Carlo simulations}
\label{subsub:DuebiSim}

\noindent In this section, we outline the Monte Carlo simulation setup used to validate the model described in \cref{sub:MCmodel}. This includes the adopted physics models, mass model extensions and scoring routines. The Monte Carlo simulations were performed to replicate the performed radiation measurements in near-, mid-, and far-field environments discussed in the previous subsections.

We employed the \texttt{precisio} physics model set in \texttt{FLUKA}, enabling high-fidelity fully coupled photon, electron, and positron transport with secondary electron production, Landau energy fluctuations, and X-ray fluorescence \citep{Battistoni2015}. To account for the non-proportional scintillation response of NaI(Tl), we implemented a non-proportional scintillation model recently published by \citet{Payne2009,Payne2011} using the user routines \texttt{comscw} and \texttt{usrglo} \citep{Breitenmoser2023,Breitenmoser2023c}. For accurate simulation of photon, electron, and positron emissions from the radionuclide sources used in the measurements described earlier, we applied the \texttt{raddecay} card in a semi-analogue mode according to the corresponding decay schemes. For the near-field measurement setup, simulations were conducted for each of the seven $\text{K}_{\text{nat}}$ sources individually, accounting for their specific activities (cf. Supplementary Table~S1). Transport thresholds were set to \qty{1}{\keV} for the scintillation crystals and surrounding components, including the reflector, optical window, and aluminum casing. Motivated by the range of the transported particles, a higher threshold of \qty{10}{\keV} was selected for the the remaining parts of the simulation domain to optimize the balance between model fidelity and computational efficiency. Additional information on the physics models, in particular the custom calibrated non-proportional scintillation model, is available in a previous work \citep{Breitenmoser2023c}. 

To account for attenuation and scattering effects, we extended the mass model described in \cref{sub:MCmodel} to encompass all major parts of the environment for each measurement scenario. Near-field simulations accounted for the hangar floor, walls, ceiling, and internal atmosphere. Mid- and far-field simulations included the surrounding soil, streets, and atmosphere. These components were modeled as homogeneous media with compositional data sourced from \citet{Mcconn2011}. The atmosphere was modeled for each measurement separately as homogeneous humid air as a function of the air temperature, pressure, and humidity. These parameters were measured by external sensors, i.e.~a HM30 meteo station for the near-field measurements and an automatic weather station (WIGOS-ID: 0-20000-0-06731) for the mid- and far-field setups. Motivated by the mean free path of the primary photons in air, atmospheric boundaries were confined to a \qty{200}{\m} radius sphere, centered at the spectrometer's location, for mid- and far-field simulations, while the hangar's dimensions defined the domain for near-field simulations. The ground was limited to a vertical extension of \qty{10}{\m} for all simulations. In addition to the environmental media, we included also detailed mass models of the deployed radionuclide sources and custom source holders. Aircraft fuel levels in each tank were determined using data from onboard fuel level probes and set accordingly in the Monte Carlo model for all simulations. For the far-field scenario, we also accurately reproduced the aircraft's position and orientation relative to the ground using system data accessed via the ARINC~429 avionics bus and flight data recorder.

As in previous works \citep{Breitenmoser2023c,Breitenmoser2022}, to simulate the spectral response of the NaI(Tl) scintillators, we scored the energy deposition events in each crystal individually on an event-by-event basis using the custom user routine \texttt{usreou} together with the \texttt{detect} card \cite{Breitenmoser2023,Breitenmoser2022a}. The number of primaries was adjusted for each simulation to ensure a median coefficient of variation \mbox{$<\!\qty{8}{\percent}$} and a relative variance of the variance \mbox{$<\!\qty{0.1}{\percent}$} over the spectral domain of interest (SDOI) in the detector channel \texttt{\#DET5}. The SDOI is defined as the spectral band between the LLD and the full energy peak (FEP) with the highest spectral energy (a more rigorous definition is provided in Supplementary Method~S1.3). Postprocessing of the obtained simulation data was performed using the \texttt{NPScinMC} pipeline detailed in Supplementary Method~S1.3. A comprehensive discussion on the uncertainty analysis, including both statistical and systematic contributions, is provided in Supplementary Method~S1.4.2.

\section{Results}
\label{sec:Results}

\subsection{Near-field scenario}
\label{sub:NearFieldResults}

\begin{figure}[tbph]
\centering
\includegraphics[width=1\textwidth]{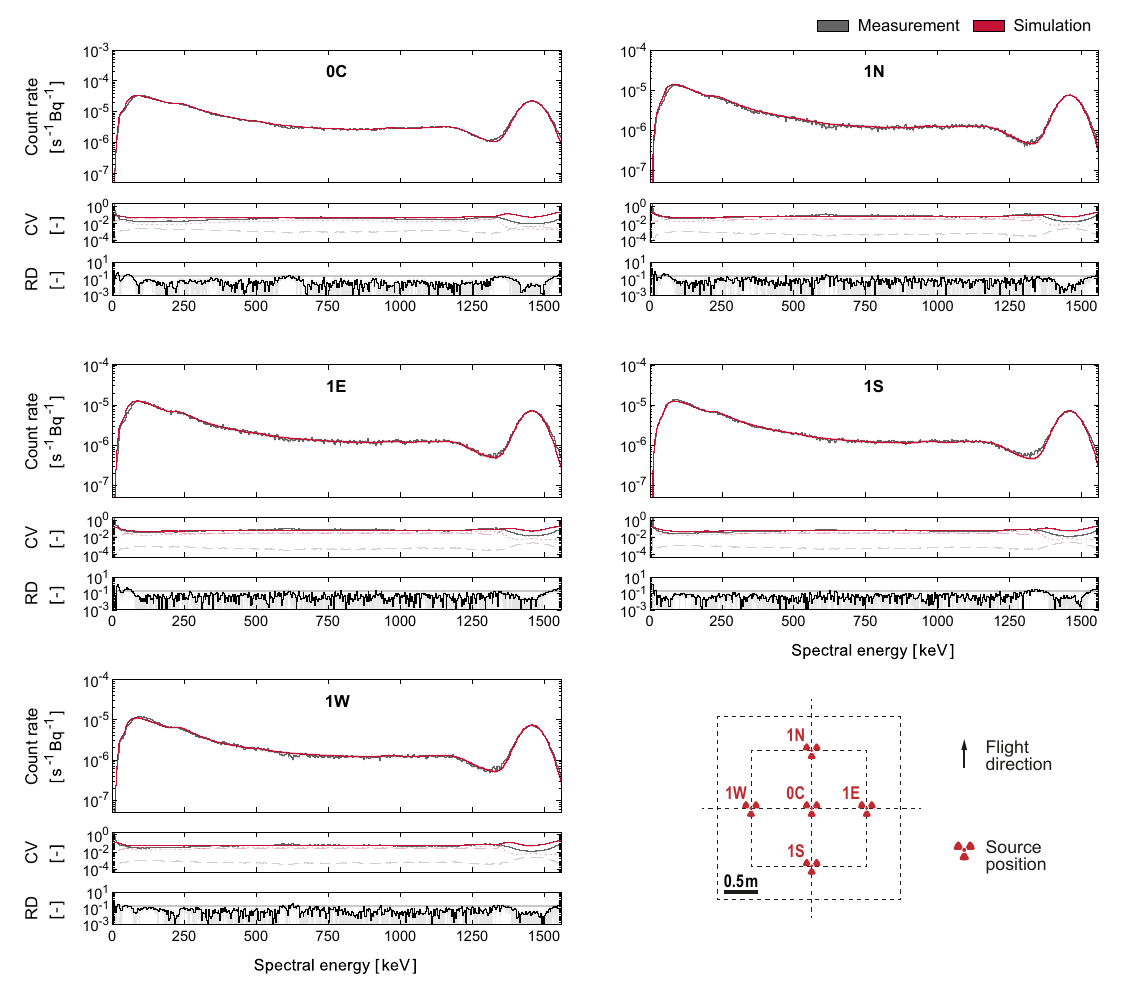}
\caption[Measured and simulated spectral signatures in the near field (0C--1W)]{Measured and simulated spectral signatures in the near field (0C--1W). The measured ($\hat{c}_{\mathrm{exp}}$) and simulated ($\hat{c}_{\mathrm{sim}}$) mean spectral signatures in the acquisition channel \texttt{\#DET5} are shown as a function of the spectral energy $E'$ with a spectral energy bin width of \mbox{$\Updelta{E'}\sim\qty{3}{\keV}$} and with the \mbox{$\text{K}_{\text{nat}}$} radionuclide volume source ($\mathcal{A} = \qty{456.6(2.6)}{\kilo\becquerel}$) placed in the near field at the predefined positions 0C, 1N, 1E, 1S, and 1W (see \cref{fig:NearFieldSetup}). Uncertainties ($\hat{\sigma}_{\mathrm{exp}}$, $\hat{\sigma}_{\mathrm{sim}}$) are provided as 1~standard deviation shaded areas. In addition, the coefficient of variation (CV) for the measured and simulated signatures as well as the relative deviation (RD) computed as $|{\hat{c}_{\mathrm{sim}}-\hat{c}_{\mathrm{exp}}}|/\hat{c}_{\mathrm{exp}}$ (\qty{20}{\percent} mark highlighted by a horizontal gray line) are provided. Statistical and systematic contributions to the CV are indicated by shaded dotted and dashed lines, respectively. For further details on the provided spectral quantities, see Supplementary Methods~S1.1--S1.4.}
\label{fig:NearFieldResultI}
\end{figure}

\begin{figure}[tbph]
\centering
\includegraphics[width=1\textwidth]{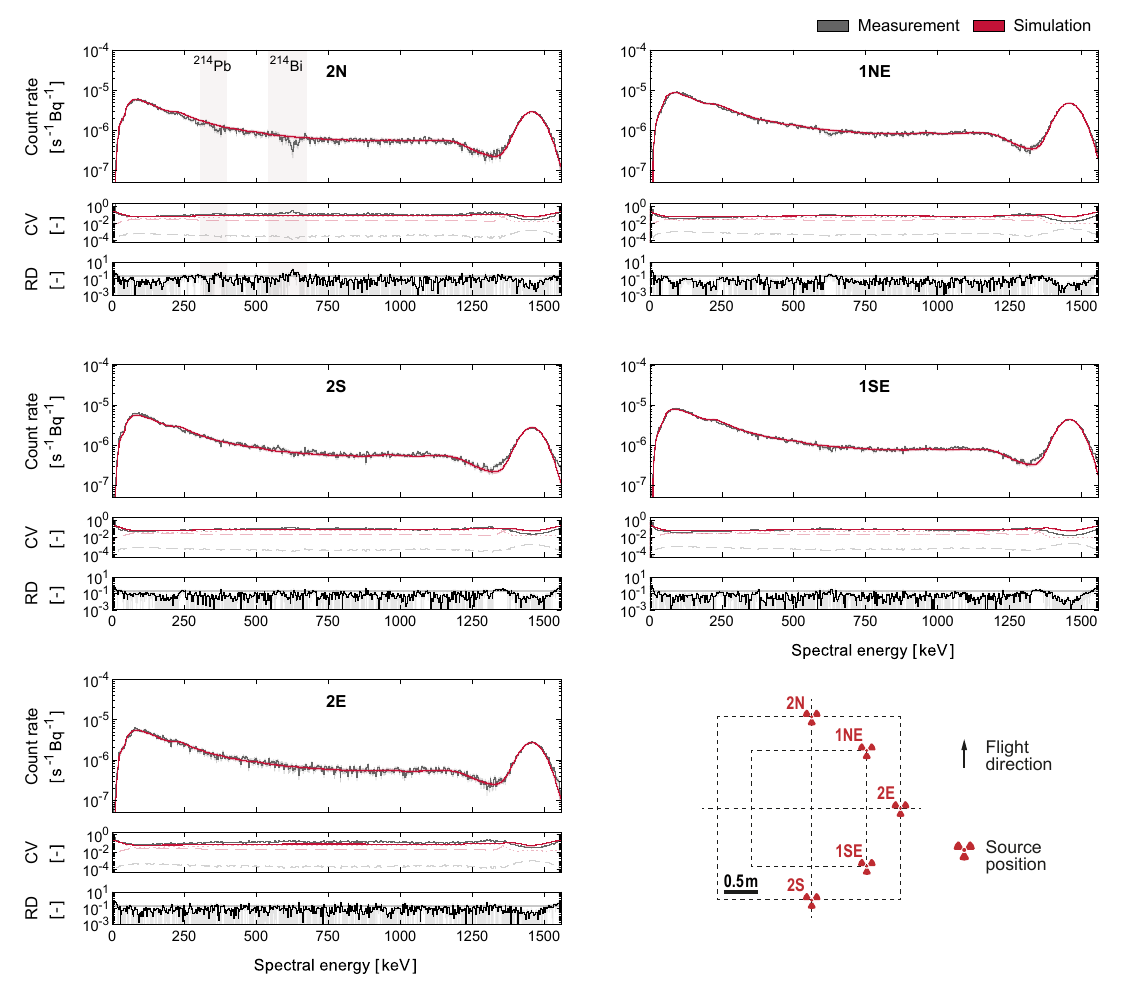}
\caption[Measured and simulated spectral signatures in the near field (1NE--2E)]{Measured and simulated spectral signatures in the near field (1NE--2E). The measured ($\hat{c}_{\mathrm{exp}}$) and simulated ($\hat{c}_{\mathrm{sim}}$) mean spectral signatures in the acquisition channel \texttt{\#DET5} are shown as a function of the spectral energy $E'$ with a spectral energy bin width of $\Updelta{E'}\sim\qty{3}{\keV}$ and with the $\text{K}_{\text{nat}}$ radionuclide volume source ($\mathcal{A} = \qty{456.6(2.6)}{\kilo\becquerel}$) placed in the near field at the predefined positions 1NE, 1SE, 2N, 2S, and 2E (see \cref{fig:NearFieldSetup}). Uncertainties ($\hat{\sigma}_{\mathrm{exp}}$, $\hat{\sigma}_{\mathrm{sim}}$) are provided as 1~standard deviation shaded areas. In addition, the coefficient of variation (CV) for the measured and simulated signatures as well as the relative deviation (RD) computed as $|{\hat{c}_{\mathrm{sim}}-\hat{c}_{\mathrm{exp}}}|/\hat{c}_{\mathrm{exp}}$ (\qty{20}{\percent} mark highlighted by a horizontal gray line) are provided. Statistical and systematic contributions to the CV are indicated by shaded dotted and dashed lines, respectively. For the source position 2N, the spectral range of the FEPs associated with the \qty{351.932(2)}{\keV} and \qty{609.312(7)}{\keV} emission lines of $^{214}\text{Pb}$ and $^{214}\text{Bi}$ is highlighted \citep{Be2008}. The spectral band widths are defined as $E_{\upgamma}\pm3\sigma_{\mathscr{\tilde{n}}}(E_{\upgamma}/\Updelta{E'})\Updelta{E'}$ with $E_{\upgamma}$, $\sigma_{\mathscr{\tilde{n}}}$, and $\Updelta{E'}$ being the photon energy of the emission line, the spectral resolution standard deviation, and the spectral energy bin width, respectively. For further details on the provided spectral quantities, see Supplementary Methods~S1.1--S1.4.}
\label{fig:NearFieldResultII}
\end{figure}

\noindent In \cref{fig:NearFieldResultI,fig:NearFieldResultII}, we present the measured ($\hat{c}_{\mathrm{exp}}$) and simulated ($\hat{c}_{\mathrm{sim}}$) spectral signatures, that are the net pulse-height count rate spectra normalized by the known source activity (see Supplementary Methods~S1.1 and S1.3), for all ten near-field source positions alongside uncertainty estimates and relative deviations computed as $|{\hat{c}_{\mathrm{sim}}-\hat{c}_{\mathrm{exp}}}|/\hat{c}_{\mathrm{exp}}$ (see Supplementary Method~S1.4). Similar to the results obtained in our previous work \citep{Breitenmoser2023c,Breitenmoser2022}, we find a good agreement between the measured and simulated spectral signatures for all ten source positions with a median relative deviation \mbox{$<\!\qty{8}{\percent}$} within the SDOI. However, there are three systematic deviations between the measured and simulated spectral signatures which require further discussion.

One of the most evident deviations is observed around the \qty{609.312(7)}{\keV} emission line of $^{214}\text{Bi}$ and, to a lesser degree, around the \qty{351.932(2)}{\keV} emission line of $^{214}\text{Pb}$ \cite{Be2008}. We highlighted the corresponding spectral range of the FEPs associated with these emission lines for the measurement 2N in \cref{fig:NearFieldResultII}, where the deviations are most pronounced. Considering that the large hangar doors were regularly opened during the field measurement campaign to allow for aircraft to be moved in and out, these deviations can be attributed to the changing background levels of the radon progeny $^{214}\text{Bi}$ and $^{214}\text{Pb}$ in the hangar atmosphere between the gross and background measurements.

A second systematic deviation can be observed at the low end of the spectral domain $E'\lesssim\qty{100}{\keV}$. Similar deviations were observed in earlier laboratory validation measurements and are attributed primarily to increased systematic uncertainties in the calibration models (spectral energy, resolution, and LLD) as well as in the physics models implemented in \texttt{FLUKA} \citep{Breitenmoser2022}.

A third notable deviation appears in the Compton gap between the full energy peak (FEP) and the Compton edge. Its consistent magnitude across all source positions aligns with the hypothesis previously proposed by \citet{Breitenmoser2023c}, suggesting deficiencies in the NPSM as the root cause.

The observed systematic deviations, except for the first one, are restricted to \mbox{$<\!\qty{20}{\percent}$} over the entire SDOI, which is well within the range of typical statistical uncertainties observed in AGRS surveys \citep{Breitenmoser2024a}. The first deviation primarily reflects long-term changes in radon progeny levels in the atmosphere evolving over the longest measurements with live times \mbox{\ensuremath{\mathcal{O}(\num{1d5})\,\unit{\second}}}, which were necessary to achieve the required measurement precision with the available sources. Since live times in AGRS are typically \mbox{\ensuremath{\mathcal{O}(\num{1d0})\,\unit{\second}}} \citep{Breitenmoser2024a}, these systematic deviations are statistically insignificant during surveys. This is also supported by the fact that the systematic deviations observed around the \qty{609.312(7)}{\keV} and \qty{351.932(2)}{\keV} emission lines tend to increase with measurement live time and are absent in the short-term measurements conducted in the mid- and far-field.

\subsection{Mid-field scenario}
\label{sub:MidFieldResults}

\begin{figure}[tbph]
\centering
\includegraphics[width=0.95\textwidth]{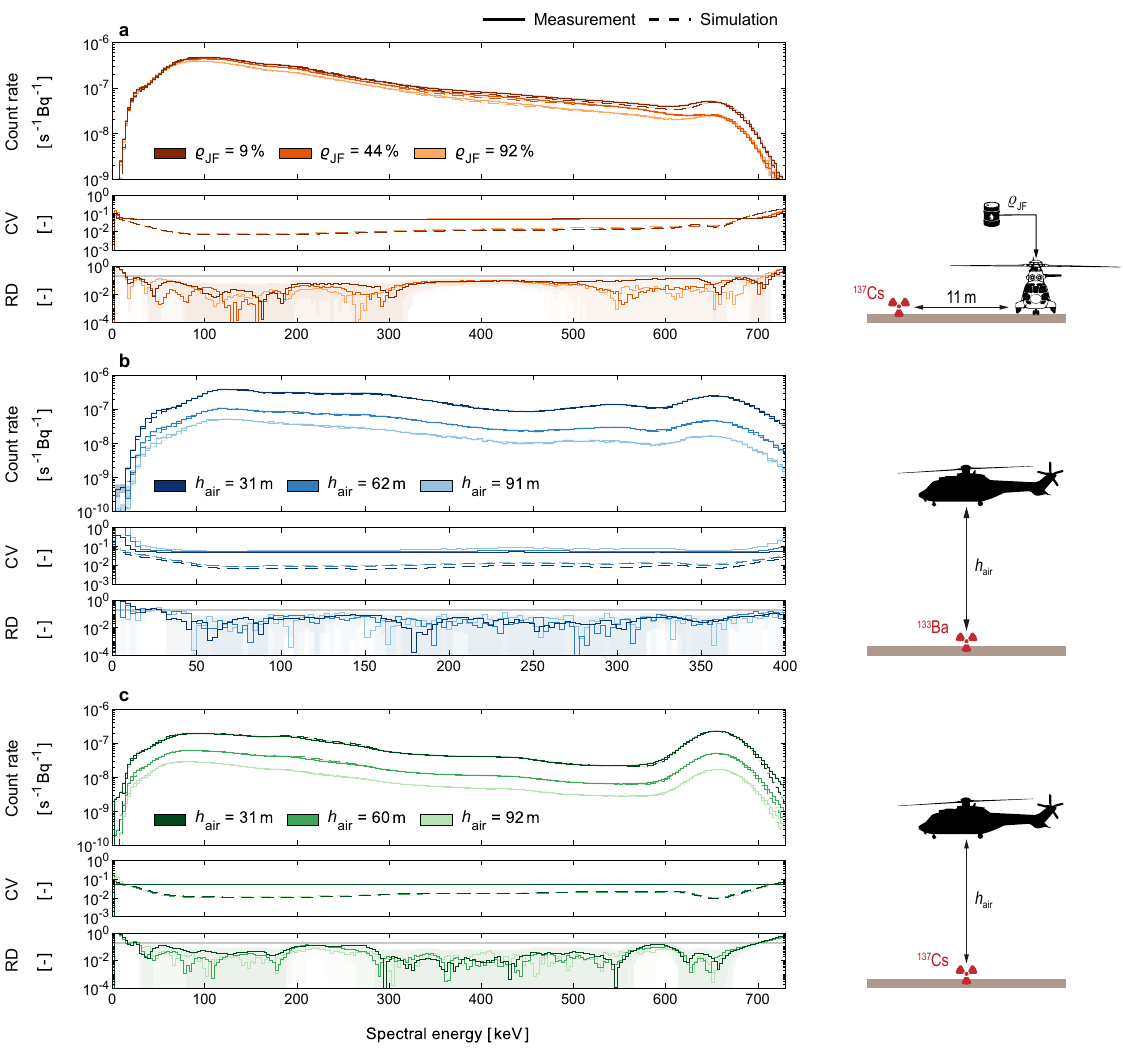}
\caption[Measured and simulated spectral signatures for the mid- and far-field scenarios]{Measured and simulated spectral signatures for the mid- and far-field scenarios. The measured ($\hat{c}_{\mathrm{exp}}$) and simulated ($\hat{c}_{\mathrm{sim}}$) mean spectral signatures in the acquisition channel \texttt{\#DET5} are shown as a function of the spectral energy $E'$ with a spectral energy bin width of $\Updelta{E'}\sim\qty{3}{\keV}$. (a)~Mid-field measurement setup with a \mbox{$^{137}\text{Cs}$} radionuclide point source ($\mathcal{A} = \qty{7.7(4)d8}{\becquerel}$) placed in a distance of \qty{11}{\m} to the starboard side of the parked AS332M1 helicopter. Measurements and simulations were repeated three times with three different fuel volume fractions ($\varrho_{\scriptscriptstyle\mathrm{JF}}$): \qty{9}{\percent}, \qty{44}{\percent}, and \qty{92}{\percent}. (b)~~Far-field measurement setup with a \mbox{$^{133}\text{Ba}$} radionuclide point source ($\mathcal{A} = \qty{4.7(2)d8}{\becquerel}$) placed on the ground below the hovering helicopter. Measurements and simulations were repeated three times at three different mean ground clearances ($h_{\mathrm{air}}$): \qty{31}{\m}, \qty{62}{\m}, and \qty{91}{\m}. (c)~~Far-field measurement setup with a \mbox{$^{137}\text{Cs}$} radionuclide point source ($\mathcal{A} = \qty{9.0(5)d9}{\becquerel}$) placed on the ground below the hovering helicopter. Measurements and simulations were repeated three times at three different mean ground clearances ($h_{\mathrm{air}}$): \qty{31}{\m}, \qty{60}{\m}, and \qty{92}{\m}. Uncertainties ($\hat{\sigma}_{\mathrm{exp}}$, $\hat{\sigma}_{\mathrm{sim}}$) are provided as 1~standard deviation shaded areas. In addition, the coefficient of variation (CV) for the measured and simulated signatures as well as the relative deviation (RD) computed as $|{\hat{c}_{\mathrm{sim}}-\hat{c}_{\mathrm{exp}}}|/\hat{c}_{\mathrm{exp}}$ (\qty{20}{\percent} mark highlighted by a horizontal gray line) are provided. Statistical and systematic contributions to the CV are indicated by shaded dotted and dashed lines, respectively. For further details on the provided spectral quantities, see Supplementary Methods~S1.1--S1.4. Line drawings of the AS332M1 helicopter were adapted from Jetijones, \href{https://creativecommons.org/licenses/by/3.0}{\texttt{CC~BY~3.0}}, via Wikimedia Commons.}
\label{fig:MidFarFieldResult}
\end{figure}

\noindent In \hyperref[fig:MidFarFieldResult]{Fig.~\ref{fig:MidFarFieldResult}(a)}, we present the measured ($\hat{c}_{\mathrm{exp}}$) and simulated ($\hat{c}_{\mathrm{sim}}$) spectral signatures obtained in the mid-field setup at three different fuel volume fractions, alongside uncertainty estimates and relative deviations computed as $|{\hat{c}_{\mathrm{sim}}-\hat{c}_{\mathrm{exp}}}|/\hat{c}_{\mathrm{exp}}$ (see Supplementary Method~S1.4). For the selected source-detector configuration, we observe a significant increase in the measured count rate of the FEP at the lowest fuel volume fraction (\qty{9}{\percent}), while minimal variation occurs between the two higher fractions. Conversely, the count rate in the Compton continuum decreases steadily with increasing fuel volume fraction.

These trends are explained by the predetermined sequence in which the fuel tanks are refilled. Tank \#2, which obstructed the line of sight for the primary photons in the adopted source-detector configuration, is among the first to be filled during refueling (see also Supplementary Fig.~S5). At a fuel volume fraction of \qty{9}{\percent}, only the lower part of tank \#2 is filled, resulting in reduced attenuation of the primary photons and a corresponding increase in the FEP count rate. At fuel volumes above \qty{\sim20}{\percent}, tank \#2 is completely filled, maximizing attenuation and stabilizing FEP count rates. In contrast, the steady decline in the Compton continuum with increasing fuel levels arises from the combined shielding effect of all six fuel tanks surrounding the spectrometer (see \cref{fig:ModelOverview}). This arrangement effectively attenuates secondary photons scattered in the environment and within the helicopter's fuselage, leading to a continuous reduction in the Compton continuum count rate as fuel levels rise.

The performed Monte Carlo simulations replicate both trends well, showing median relative deviations of \qty{7.5}{\percent}, \qty{3.5}{\percent}, and \qty{2.8}{\percent} within the SDOI for the fuel volume fractions of \qty{9}{\percent}, \qty{44}{\percent}, and \qty{92}{\percent}, respectively. However, we observe a systematic underestimation in the spectral signature by our Monte Carlo model starting at the lower end of the backscatter peak at \mbox{$E'\sim\qty{200}{\keV}$} and extending up to the FEP for all three fuel volume fractions. The magnitude of the relative deviation increases as the fuel volume fraction decreases, suggesting that this deviation is likely due to systematic uncertainties in the fuselage mass model. Specifically, the approximation of the AS332M1 aircraft's complex semi-monocoque fuselage design to a homogeneous equivalent metal sheet is expected to introduce a systematic error in the spectral detector response. In contrast, the aircraft fuel, being a liquid, is well approximated as a homogeneous mass model element. Consequently, as the fuel level decreases, the relative influence of the fuel diminishes while the effect of the fuselage becomes more pronounced, leading to increased deviations in the Compton continuum, Compton gap, and the FEP. It is worth adding that the relative deviation is again restricted to \mbox{$<\!\qty{20}{\percent}$} over the entire SDOI, even for the lowest evaluated fuel volume fraction.

\subsection{Far-field scenario}
\label{sub:FarFieldResults}

\noindent In \hyperref[fig:MidFarFieldResult]{Fig.~\ref{fig:MidFarFieldResult}}, we present the measured ($\hat{c}_{\mathrm{exp}}$) and simulated ($\hat{c}_{\mathrm{sim}}$) spectral signatures obtained in the far-field source-detector configuration for the two deployed radionuclide point sources, i.e.~\mbox{$^{133}\text{Ba}$} in \hyperref[fig:MidFarFieldResult]{Fig.~\ref{fig:MidFarFieldResult}(b)} and \mbox{$^{137}\text{Cs}$} in \hyperref[fig:MidFarFieldResult]{Fig.~\ref{fig:MidFarFieldResult}(c)}, alongside uncertainty estimates and relative deviations computed as $|{\hat{c}_{\mathrm{sim}}-\hat{c}_{\mathrm{exp}}}|/\hat{c}_{\mathrm{exp}}$ (see Supplementary Method~S1.4).

As predicted by the monoenergetic transport theory \citep{Kogan1971,Breitenmoser2024a}, we find an exponential decrease in the count rate in the FEPs with increasing ground clearance. In general, good agreement between the measured and simulated spectral signatures is observed for both radionuclides at all three ground clearances with a median relative deviation \mbox{$<\!\qty{5}{\percent}$} within the SDOI. However, similar to the near- and mid-field scenarios, there are also some systematic deviations between the measured and simulated spectral signatures which require further discussion.

One of the most evident deviations is observed at low spectral energies in the backscatter peak for the \mbox{$^{137}\text{Cs}$} source and, to a lesser degree, around \qty{\sim150}{\keV} for the \mbox{$^{133}\text{Ba}$} source. The fact that the magnitude of the relative deviation tends to decrease with increasing ground clearance indicates that it is likely related to systematic uncertainties in the mass model media close to the ground, i.e.~the source holder components, tarmac and soil. A sensitivity analysis confirmed the backscatter peak's strong dependence on variations in these environmental components (see Supplementary Fig.~S7).

A second deviation is found in the Compton gap between the Compton edge and the FEP for the \mbox{$^{137}\text{Cs}$} source. Similar deviations were noted already in our validation measurements under laboratory conditions \citep{Breitenmoser2023c}. As with the deviations observed at the backscatter peak, the magnitude of the relative deviation tends to decrease with increasing ground clearance. Consequently, these deviations may be attributed to systematic uncertainties in the mass model media close to the ground, too. Alternatively, as proposed by \citet{Breitenmoser2023c}, these deviations could also stem, at least partly, from deficiencies in the adopted NPSM. It is worth noting that the observed relative deviations are again restricted to \mbox{$<\!\qty{20}{\percent}$} over the entire SDOI for all evaluated hover flights, which is well within the range of typical statistical uncertainties observed in AGRS surveys \citep{Breitenmoser2024a}.

\section{Conclusion}
\label{sec:Conclusion}

Here, we developed a high-fidelity Monte Carlo model to enable full-spectrum calibration of the Swiss AGRS system. Unlike previous approaches \citep{Allyson1998,Billings1999,Sinclair2011,Torii2013,Sinclair2016,Zhang2018,Kulisek2018}, our model integrates a detailed mass model of the entire aircraft and an advanced non-proportional scintillation model, allowing for accurate event-by-event simulations of the NaI(Tl) crystals' scintillation response to arbitrarily complex gamma-ray fields. Moreover, rather than restricting the modeling to narrow spectral bands, as done in most previous work \citep{Billings1999,Sinclair2011,Torii2013,Sinclair2016,Zhang2018}, our Monte Carlo model reproduces the scintillation response across the full spectral range of the spectrometer, spanning from \qty{\sim50}{\keV} to \qty{\sim3}{\MeV}.

Extensive validation measurements in near-, mid-, and far-field source-detector configurations with various radionuclide sources demonstrated that the developed Monte Carlo model not only effectively addresses the large deficiencies of previous models, reducing relative errors by a factor \mbox{\ensuremath{\mathcal{O}(\num{1d1})}}, but also achieves, for the first time, the accuracy and precision required to supersede traditional empirical calibration methods. Residual systematic deviations remain in general small and are well within the range of typical statistical uncertainties observed in AGRS surveys \citep{Breitenmoser2024a}, ensuring the model’s reliability for practical applications. 

These results mark a major advancement in AGRS. Unlike existing empirical methods \cite{IAEA2003,IAEA1991}, the developed Monte Carlo model enables high-fidelity, full-spectrum calibration of any terrestrial radionuclide source, particularly those emitting low-energy photons (\mbox{$\lesssim\!\qty{400}{\keV}$}) essential for emergency response to radiological incidents and homeland security applications. Moreover, Monte Carlo based calibration offers substantial time and cost savings, minimizes reliance on high-intensity calibration sources, and eliminates the generation of related radioactive waste from long-lived radionuclide calibration sources, including calibration pads traditionally used in AGRS \citep{IAEA2003,IAEA1991,Dickson1981,Grasty1991,Minty1990}. 

It is important to note that during near-field measurements with the longest live times \mbox{\ensuremath{\mathcal{O}(\num{1d5})\,\unit{\second}}}, enhanced systematic deviations were observed around the \qty{609.312(7)}{\keV} emission line of $^{214}\text{Bi}$ and, to a lesser degree, around the \qty{351.932(2)}{\keV} emission line of $^{214}\text{Pb}$, attributed to changes in atmospheric radon progeny levels. Given that AGRS surveys are conducted with much shorter live times \mbox{\ensuremath{\mathcal{O}(\num{1d0})\,\unit{\second}}}, these deviations are statistically insignificant for AGRS applications, as confirmed by the absence of such deviations in the shorter measurements conducted in the mid- and far-field. The observed discrepancies exemplify the background biases inherent in long-term radiation measurements under field conditions \citep{Amestoy2021,Baldoncini2018}. Since empirical calibration methods rely on such long-term measurements \citep{IAEA2003,IAEA1991,Dickson1981,Grasty1991,Minty1990}, the presented results do not undermine but rather highlight the advantage of numerical calibration methods, which are free of any background biases. Additionally, owing to its detailed design over the full \mbox{\ensuremath{4\uppi}} solid angle, the developed Monte Carlo model not only extends the calibration range to any terrestrial gamma-ray emitting radionuclide, but also enables the generation of high-fidelity spectral signatures for atmospheric gamma-ray sources. This opens exciting opportunities for AGRS, including real-time background correction and novel operational capabilities, such as probing trace airborne radionuclides emitted by nuclear facilities to support the verification of nuclear treaty compliance or quantifying the cosmic-ray flux and radon progeny activity concentrations in the lower atmosphere for advancing geophysical research.

Although this work focused on the Swiss AGRS system, given the similarities in gamma-ray spectrometers and aircraft platforms \citep{Breitenmoser2024a}, the developed Monte Carlo based full-spectrum modeling approach is broadly applicable to other AGRS systems worldwide. The main challenge in transferring the developed calibration methodology to other systems lies in the considerable effort required to develop the Monte Carlo models for the specific system configurations and source-detector scenarios encountered in practice. A particularly demanding aspect is the creation of detailed aircraft mass models, requiring extensive material and geometric data. To address these challenges, future efforts could explore the integration of CAD-based software tools or the development of generic aircraft mass models to streamline the modeling process \citep{Haußler2017,Wilson2010,Weinhorst2015,Davis2019}. Such advancements would not only reduce development time but also enhance the scalability and accessibility of the methodology for diverse applications. 

The successful demonstration of high-fidelity, full-spectrum calibration for AGRS systems is a critical prerequisite for advancing full-spectrum analysis (FSA) pipelines by enabling the unrestricted generation of accurate spectral signature estimates for these systems \citep{Grasty1985,Minty1998a,Hendriks2001,Paradis2020,Andre2021}. Previous studies suggest that FSA techniques significantly enhance both sensitivity by at least \qty{300}{\percent} and accuracy by \qty{25}{\percent} compared to traditional spectral window and peak-fitting approaches \citep{Grasty1985,Hendriks2001,Caciolli2012}. Building on these findings, the Monte Carlo based full-spectrum calibration approach is expected to systematically refine source identification, quantification, and localization by AGRS, with broad implications for enhancing the capabilities of these systems in environmental monitoring, emergency response, and nuclear security.

While the Monte Carlo based full-spectrum modeling approach for AGRS systems has proven superior to traditional empirical calibration methods in the field scenarios described in this study, its feasibility for integration into FSA pipelines has yet to be demonstrated, with associated computational cost posing the primary challenge. To achieve the required precision, a Monte Carlo simulation of a single spectral signature requires a characteristic computation time of {\ensuremath{\Updelta{t}_{\mathrm{MC}}=\mathcal{O}(\num{1d4})\,\unit{\corehour}}} on a typical computer cluster, such as the one used in this study. Moreover, due to the inherent variability in the source-detector configurations as well as the various changes in the atmospheric properties and the terrestrial scene during a survey flight, spectral signatures need to be predicted for each individual recorded pulse-height spectrum and each gamma-ray source of interest. As a result, the number of model evaluations necessary for the analysis of a typical AGRS survey flight with $N_{\mathrm{spec}}=\mathcal{O}(\num{1d4})$ recorded spectra and $N_{\mathrm{src}}=\mathcal{O}(\num{1d1})$ gamma-ray sources of interest is in the order of $N_{\mathrm{src}}N_{\mathrm{spec}}\Updelta{t}_{\mathrm{MC}}=\mathcal{O}(\num{1d9})\,\unit{\corehour}$. 

Given these numbers, it is evident that calibrating AGRS systems for extended AGRS surveys using brute-force Monte Carlo simulations becomes computationally prohibitive with current cluster infrastructure, and by extension, any application in spectral data reduction tasks such as quantification, identification, or localization. This limitation of Monte Carlo simulations is nevertheless not unique to AGRS systems but is a recurring challenge across numerous scientific disciplines where radiation spectrometers with complex spectral and angular dispersion characteristics are operated in varying source-detector configurations. In astrophysics \citep{Ackermann2012,Kaastra2016,Duan2019,Luo2020,Chen2013} and planetary science \citep{Reedy1973,Lawrence2010a,Prettyman2011}, surrogate modeling techniques that leverage Monte Carlo derived detector response functions have proven effective in generating accurate, near real-time spectral predictions. Future work is necessary to evaluate the potential of this surrogate approach in addressing the computational challenges of Monte Carlo based full-spectrum modeling for AGRS applications.

\bibliographystyle{elsarticle-num-names} 

\section*{CRediT authorship contribution statement}

\noindent\textbf{D.~Breitenmoser:} Conceptualization, Data curation, Formal analysis, Investigation, Methodology, Project administration, Software, Supervision, Validation, Visualization, Writing -- original draft, Writing -- review and editing. \textbf{A.~Stabilini:} Investigation, Writing -- review and editing. \textbf{M.~M.~Kasprzak:} Writing -- review and editing. \textbf{S.~Mayer:} Funding acquisition, Writing -- review and editing.

\section*{Declaration of competing interest}

\noindent The authors declare that they have no known competing financial interests or personal relationships that could have appeared to influence the work reported in this paper.

\section*{Acknowledgments}

\noindent We gratefully acknowledge the support by the members of the National Emergency Operations Centre (NEOC), the Swiss Armed Forces, specifically the Swiss Air Force and the Nuclear, Biological, Chemical, Explosive Ordnance Disposal and Mine Action Centre of Competence (NBC-EOD), as well as the Expert Group Airborne Gamma Spectrometry for their support in conducting the validation measurements described in this study. Our sincere thanks go to Gernot Butterweck for his invaluable scientific expertise and support during the measurements, as well as for his role in supervision. We also thank Eduardo Gardenali Yukihara and Federico Alejandro Geser for their valuable input during the internal review phase of the manuscript. Finally, we extend our gratitude to Dominik Werthmüller for his technical support in running the Monte Carlo simulations on the computer cluster at the Paul Scherrer Institute. This research was partially supported by the Swiss Federal Nuclear Safety Inspectorate (grant no. CTR00836 \& CTR00491).

\section*{Data and code availability}

\noindent The \texttt{FLUKA} code \citep{Ahdida2022} used for Monte Carlo simulations is available at \url{https://fluka.cern/}. We adopted the graphical user interphase \texttt{FLAIR} \citep{Vlachoudis2009}, available at \url{https://flair.web.cern.ch/flair/}, to setup the \texttt{FLUKA} input files and create the mass model figures. The custom \texttt{FLUKA} user routines employed in the Monte Carlo simulations have been deposited on the ETH Research Collection repository under accession code \url{https://doi.org/10.3929/ethz-b-000595727} \citep{Breitenmoser2023} and \url{https://doi.org/10.3929/ethz-b-000528892} \citep{Breitenmoser2022a}. Data processing and figure creation was performed by the $\texttt{MATLAB}\textsuperscript{\textregistered}$ code. Measurement data will be made available on request.

\end{document}